\documentclass[conference]{IEEEtran}
\IEEEoverridecommandlockouts
\usepackage{cite}
\usepackage{amsmath,amssymb,amsfonts,mathtools}
\usepackage{amsthm}
\usepackage{algorithmic}
\usepackage{bbm}
\usepackage{graphicx}
\usepackage{textcomp}
\usepackage{xcolor}

\allowdisplaybreaks

\renewcommand{\baselinestretch}{0.94}
\def\BibTeX{{\rm B\kern-.05em{\sc i\kern-.025em b}\kern-.08em
    T\kern-.1667em\lower.7ex\hbox{E}\kern-.125emX}}
\begin{document}

\title{\huge Reinforcement Learning-Enabled Reliable Wireless Sensor Networks in Dynamic Underground Environments
}

\author{\IEEEauthorblockN{ Hongzhi Guo and Bincy Ben } 
	\IEEEauthorblockA{ Engineering Department, Norfolk State University\\
		Email: hguo@nsu.edu, b.ben@spartans.nsu.edu\\
}
}

\maketitle

\begin{abstract}
Wireless underground sensor networks play an important role in underground sensing such as climate-smart agriculture and underground infrastructure monitoring. Existing works consider a static underground environment, which is not practical since the dielectric parameters of soil change frequently due to precipitation and harsh weather. This challenge cannot be ignored in real implementation due to the drastic change of wireless underground channel. In this paper, we study the effect of dynamic underground environment on wireless communications for sensor networks. We use the real data collected by in-situ sensors to train a Hidden Markov Model. Then, by using reinforcement learning, we derive the optimal transmission policies for underground sensors to efficiently use their energy and reduce the number of dropped and unsuccessfully transmitted packets. Through simulations using real data, we find that the developed algorithm can reduce the packet loss and transmit the sensed data in a timely manner.   
\end{abstract}


\section{Introduction}
The recent development of the Internet of Things (IoT) promises to connect billions of smart devices. The dynamic underground environment is a frontier of IoT, where wireless sensors are employed to monitor soil status for designing climate-smart agriculture systems, studying forest dynamics, and detecting underground pipeline leakage \cite{vuran2018internet}. Different from remote sensing or ground penetration radar, underground sensors have unprecedented advantages in providing accurate real-time in-situ sensing.  

The main challenge of wireless underground communications resides in penetrating through the dense and lossy soil medium, due to which the communication range is very limited and the power consumption is huge. Wireless channels of electromagnetic (EM) waves \cite{dong2013environment} and magnetic induction (MI) \cite{guo2016increasing} have been extensively studied for underground communications. The carrier frequency of EM waves are several hundred MHz and, thus, the data rate is high, while the MI uses low frequency to create long wavelength to reduce the propagation loss, which results in low data rate. Although designing an efficient wireless network in lossy soil medium is already challenging, its dynamic change further increases the complexity.

Existing works consider a static underground environment, where dielectric parameters of soil do not change with time. In reality, due to precipitation, the water content of soil changes frequently. Since the relative permittivity of water is around 81, which is much larger than the dry soil's (around 2), the water content significantly affects the soil's dielectric parameters. The same effect can be observed for electric conductivity. Although the works in \cite{vuran2018internet,dong2013environment,guo2015text} consider various water content and conductivity of soil, there is no discussion on the continuous change of soil dielectric parameters. On one hand, the high conductivity of soil reduces the communication range, while the low conductivity of soil allows long-range communication. If we design wireless underground sensor networks by considering the worst case, i.e., the highest conductivity, the lifetime of networks can be very short due to the high power consumption. On the other hand, if the design considers the best case, i.e., lowest conductivity, the connectivity and packet loss can be dramatically increased when the precipitation is high, which increase the soil conductivity in a short period. Therefore, we need optimal communication policies for wireless sensors in dynamic underground environments, which can be adaptive to environmental changes.

Reinforcement learning (RL) is an effective approach to solve dynamic problems. The dynamic environment is modeled as Hidden Markov Models and optimal actions can be derived using Bellman's equation. In \cite{ku2015data}, RL is used for energy harvesting networks to optimal use of the harvested energy to communicate. In \cite{li2018reinforcement}, RL is used to schedule wireless power transfer for wireless sensors based on their battery level. In \cite{wang2018deep}, deep RL is used for dynamic multichannel access in wireless networks. Although not listed here, RL has shown its unprecedented advantages in dealing with dynamic changes in wireless communications.   

In this paper, we propose a data-driven model to capture the dynamic change of dielectric parameters of soil, upon which we develop an adaptive model to efficiently use sensor's energy and reduce the number of packet loss. We use the underground soil dielectric data from Nevada Climate Change Portal \cite{dascalu2014overview}, which is collected by underground sensors at Snake Range West Montane in 2017, to develop a Markov Decision Process to capture the dynamic change. From the data, we notice that some measurements are missing, which not only shows the low reliability of the system but also motivates this work. We study the impact of the soil parameter change on wireless channels and derive states based on the change of channel path loss. Then, we use RL to obtain the optimal transmission policies at each state. We show that the delay and the packet loss are controllable and the system is more reliable and efficient than transmission policies considering static environments. To the best of our knowledge, this is the first paper that investigates the optimal transmission policies for wireless underground sensor networks considering the dynamic change of the soil medium.      

The rest of this paper is organized as follows. In Section II, we introduce the system model, including the dynamic environment, channel, states, actions, and the reward model. After that, we derive the optimal policies by using reinforcement learning in Section III. In Section IV, we present the numerical simulations and insights on the optimal policies. Finally, this paper is concluded in Section V.  

\section{System Model}
In this section, we present the system model to capture key factors that can affect the system performance. First, we introduce the communication protocol to support the developed optimal transmission policy. 
\begin{figure}[t]
	\centering
	\includegraphics[width=0.25\textwidth]{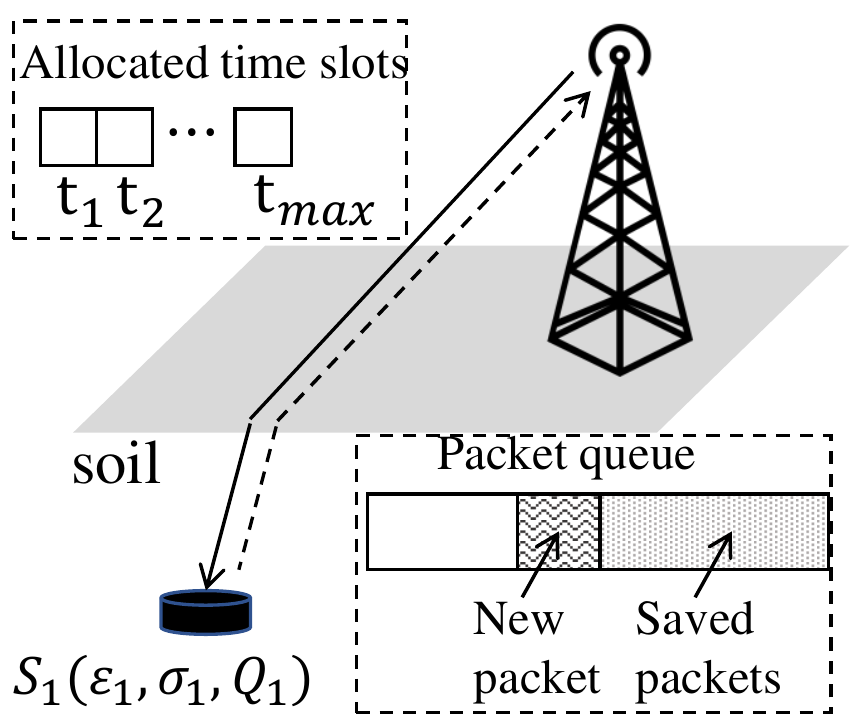}
	\vspace{-5pt}
	\caption{Illustration of the proposed wireless underground sensor network considering dynamic environmental changes.}
	\vspace{-10pt}
	\label{fig:sys}
\end{figure}
\subsection{Communication Protocol}
In the underground sensor networks, there are multiple sensors buried in soil. In this paper, we focus on the performance of a single wireless sensor to show the improvement of the proposed adaptive transmission policy. We consider that the sensor $S_1$ is buried underground with depth $d_{ug}$. The soil's permittivity is $\epsilon_1$ and conductivity is $\sigma_1$. The sensor also has a packet queue, which can save packets when the channel status is not good. The number of packets in the queue is denoted by $Q_1$ and $0 \leq Q_1\leq N_q $. The data packet is generated periodically since the sensor samples the environmental parameters in a periodical way (e.g., every one or ten minutes). Also, the sensor directly communicates with a basestation (BS) without rerouting packets. Note that, $\epsilon_1$, $\sigma_1$, and $Q_i$ are time-dependent.   

In this paper, we consider framed transmission. The sensing and transmission period is  $T$ and it is divided into time slots. Since the communication takes much shorter time than the period $T$, e.g, less than 1 second compared with 10 minutes. In each time slot only one packet can be transmitted. The sensor is allocated $t_{max}$ time slots and it can decide the number of transmitted packets, which ranges from 0 (no transmission) to $t_{max}$ packets. However, due to the harsh channel and limited transmission power, not all the transmissions are successful. If the sensor does not receive an acknowledgment from BS, it saves the packet in its queue, which will be retransmitted in the next period. An illustration of the protocol is shown in Fig.~\ref{fig:sys}. Since the BS is above ground and it only sends limited amount of data to underground sensors, we focus on the uplink communication, through which sensors report sensed data to the BS.

\subsection{Effect of $\epsilon_1$ and $\sigma_1$}
The underground wireless channel is not static due to the dynamic change of underground soil conductivity and permittivity. According to \cite{dong2013environment}, considering the soil-air interface the uplink path loss between sensors and BS can be written as
\begin{align}
\label{equ:received_power}
-10\log\frac{{\tilde P}_r}{P_t}=-10\log\frac{G_tG_r}{L_{ug}(d_{ug})L_{ag}(d_{ag})L(R)},
\end{align}
where ${\tilde P}_r$ is the received power, $P_t$ is the transmission power and $L_{ug}(d_{ug})$, $L_{ag}(d_{ag})$, and $L(R)$ are the underground propagation loss, aboveground propagation loss, and reflection loss on the boundary, respectively, which are given in \cite{dong2013environment,vuran2010channel}. $L_{ug}(d_{ug})$ and $L(R)$ are functions of the propagation constant of soil, which is 
\begin{align}
k_s= j2 \pi f \sqrt{\mu_0 \left(\epsilon_1-j\frac{\sigma_1} {2\pi f}\right)}
\end{align}
where $j=\sqrt{-1}$, $f$ is the carrier frequency, $\mu_0$ is the permeability, $\epsilon_s$ is the permittivity of the soil, and $\sigma_s$ is the conductivity of the soil. Although \eqref{equ:received_power} captures the dominant components of the received power, the noise is neglected. Here, we use a more comprehensive model and the received power can be updated as $P_r={\tilde P}_r+ \eta$, where $\eta$ is the noise power.

\begin{figure}[t]
	\centering
	\includegraphics[width=0.4\textwidth]{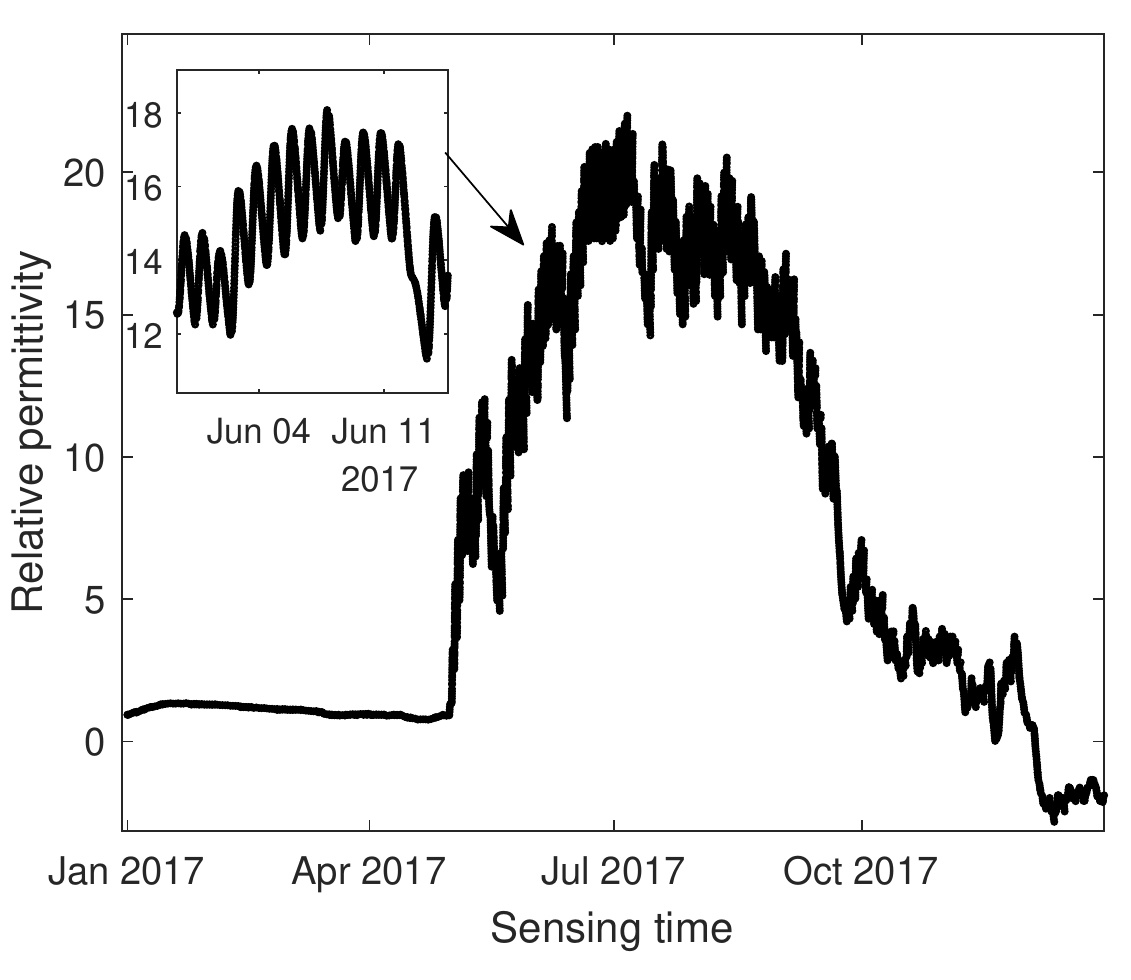}
	\vspace{-5pt}
	\caption{Soil permittivity at Snake Range West Montane in 2017. Data are downloaded from Nevada Climate Change Portal \cite{dascalu2014overview}.}
	\vspace{-10pt}
	\label{fig:snake_permittivity}
\end{figure}

\begin{figure}[t]
	\centering
	\includegraphics[width=0.35\textwidth]{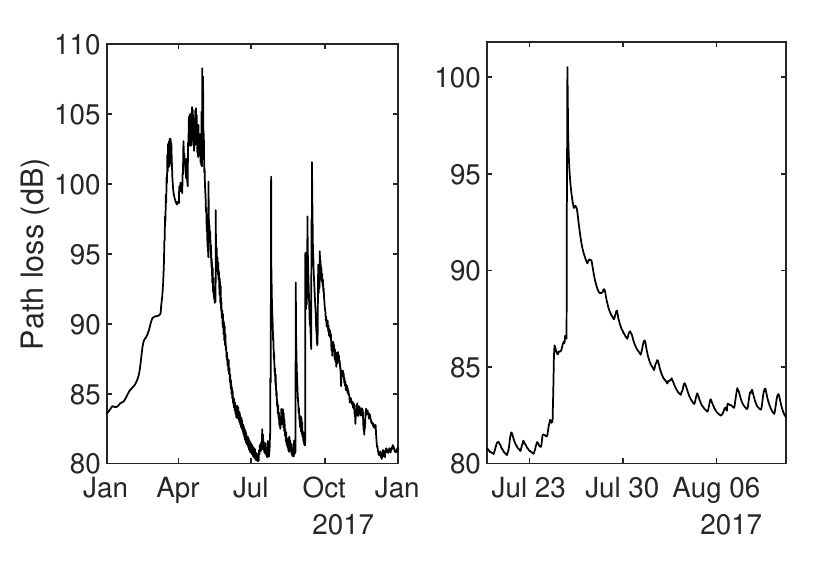}
	\vspace{-5pt}
	\caption{Dynamic path loss in 2017 (left) and zoom-in in July and August (right). $d=9.5$~cm, $f=300$~MHz, $d_{ag}=20$~m, $G_t=G_r=5$~dB.}
	\vspace{-10pt}
	\label{fig:pathloss}
\end{figure}
The received power is affected by $\epsilon_1$ and $\sigma_1$, which varies with time. In Fig. \ref{fig:snake_permittivity}, we show the soil permittivity measured every 10 minutes at Snake Range West Montane in 2017 with depth 9.5~cm. The data are collected from Nevada Climate Change Portal \cite{dascalu2014overview} which is based on a real-time underground sensing platform at multiple locations in the state of Nevada. The data show that $\epsilon_1$ changes significantly during a year, e.g., the value in summer is much larger than that in winter. Moreover, as shown in Fig.~\ref{fig:snake_permittivity}, the day-to-day change is also considerable. Although not shown here, the soil conductivity also demonstrates drastically change during the year and it also has a day-to-day changing pattern. Also, we notice the change of conductivity and permittivity are not correlated. Since the wireless channel between underground sensors and above ground BS is a function of $\epsilon_1$ and $\sigma_1$, we can expect a significant change of the channel. 

In Fig.~\ref{fig:pathloss}, the change of path loss by considering dynamic $\epsilon_1$ and $\sigma_1$ using \eqref{equ:received_power} is shown. As we can see, the path loss can change as high as 25~dB and also there are small daily changes. Moreover, the path loss may increase significantly during a couple of days. If we consider a static environment, such a large change of path loss can be ignored, which is too optimistic. Next, we train a Hidden Markov Model to capture the dynamic change of path loss.

\subsection{State Model}
Our state model consists of two parts, namely, the dynamic wireless channel due to environment changes and the sensor's packet queue. 
\subsubsection{Wireless Channel}  
Instead of modeling the dynamic change of $\epsilon_1$ and $\sigma_1$ individually, we consider their effects jointly by looking at the wireless channel model in \eqref{equ:received_power}. By substituting the continuous time series $\epsilon_1$ and $\sigma_1$ into \eqref{equ:received_power}, we obtain the channel path loss. Then, we employ the Hidden Markov Model and Gaussian emission model \cite{ku2015data} to derive the states for Markov Decision Process. Assume that there are $N_g$ Gaussian distributions ${\mathcal {N}}_n(u_n, \Sigma_n),~n=1,2,\cdots, N_g$ and $N_g$ states. Each distribution is associated with a state. Each path loss measured at the sampling time is a variable generated from a state with the associated Gaussian distribution. In this way, a time-series of the path loss is captured by using the finite state Markov model and the parameter of Gaussian distributions can be obtained by using the Expectation-Maximization algorithm. Note that in Fig.~\ref{fig:snake_permittivity}, some of the measured $\epsilon_1$ are negative which is mainly due to the sensor's low accuracy when the temperature is low. To avoid misleading results, we consider all the negative $\epsilon_1$ as 1 which can reflect the physics better. Also, due to unreliable wireless transmission, some of data are missing. Although the portion is very small, this can affect the training result. We use linear filling to obtain substitutions for the missing data. 

We consider that there are 15 states, which can reflect the change well. Although using a model with more states can obtain the optimal policy that can improve the system performance, it also increases the computation burden. In addition, we notice that simply train the states by using path loss values cannot fully capture the dynamics. For example, two samples may have the same path loss, but for the first one the next sample is increasing, while for the second one the next sample is decreasing. For the optimal transmission, the sensor should transmit more in the first case since the path loss is increasing, while for the second case the sensor should save packets if it has space since the path loss is decreasing and if it transmits in the future, rather than now, it can have better probability to successfully transmit packets. Thus, in the training we also consider the change of path loss. For each training input, it has the path loss, as well as the difference between the current path loss and the previous one. In this way, depends on the path loss is increasing or decreasing, the sensor may make different decisions.

\begin{figure}[t]
	\centering
	\includegraphics[width=0.4\textwidth]{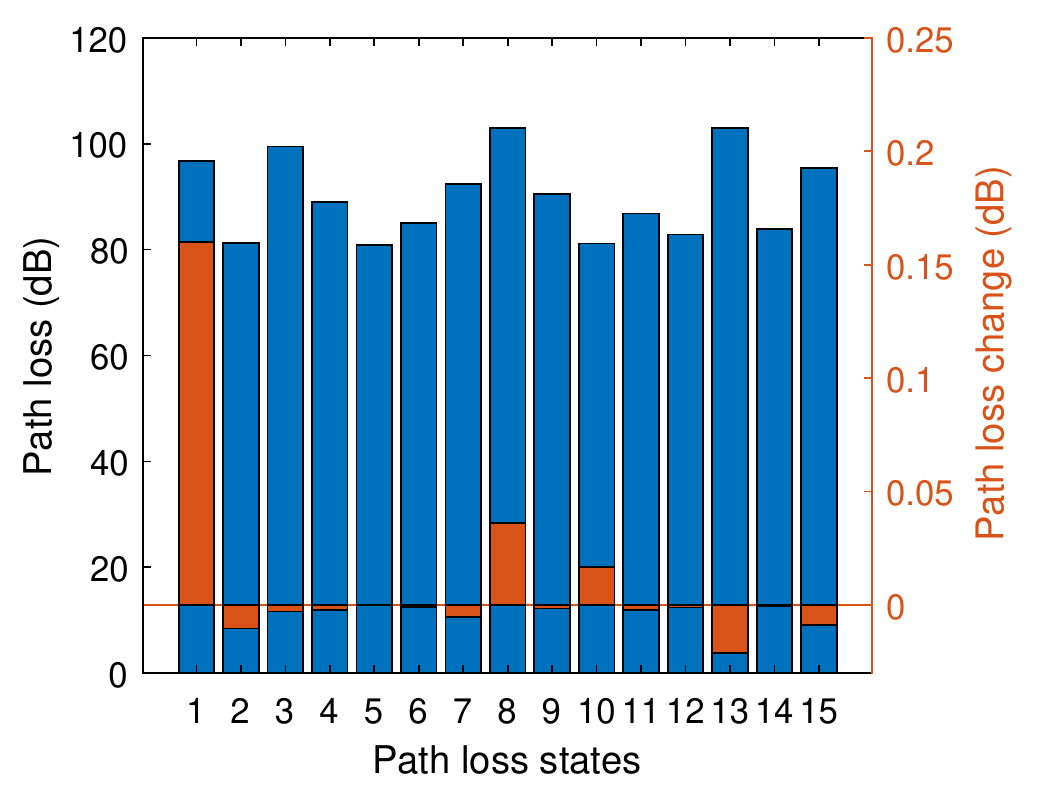}
	\vspace{-5pt}
	\caption{Mean path loss and change of path loss for each state.}
	\vspace{-10pt}
	\label{fig:state_pathloss}
\end{figure}

\begin{figure}[t]
	\centering
	\includegraphics[width=0.4\textwidth]{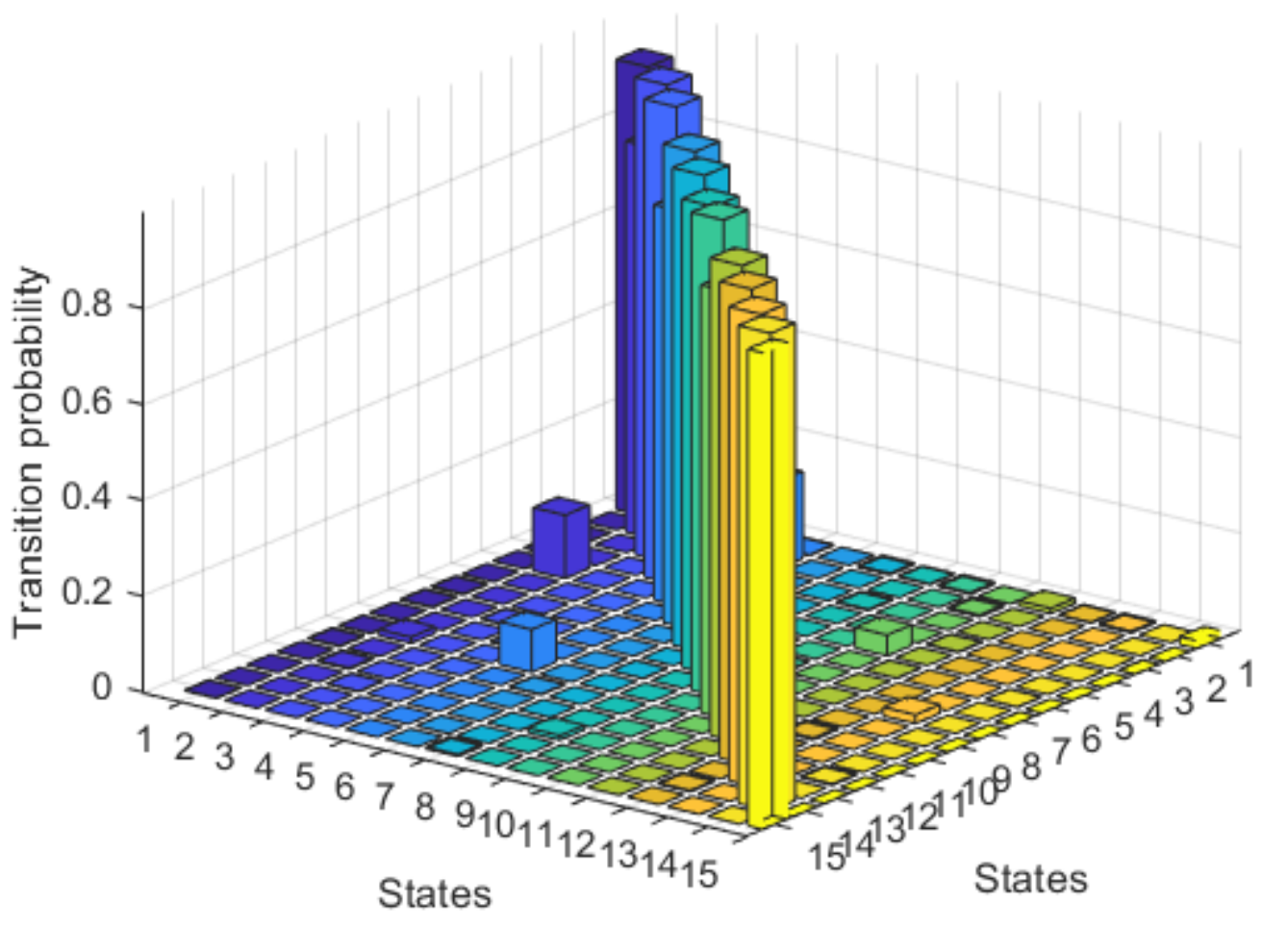}
	\vspace{-5pt}
	\caption{Transition probability of states.}
	\vspace{-10pt}
	\label{fig:transition_prob}
\end{figure}

In Fig.~\ref{fig:state_pathloss}, the mean value and change of path loss for each state is shown. As we can see from the figure, the change of path loss between two samples is small and most are around 0. The transition probability is given in Fig.~\ref{fig:transition_prob}. The probability of changing from one state to another is low and most of the times the path loss remains in its current state. The probability of state transition can be written as $P(c'|c)$, where $c$ is the current channel path loss and $c'$ is the next channel path loss. 


\subsubsection{Packet Queue}
The sensor measures the information of interest every $T$ and generate a data packet. Traditional solution is transmitting the packet immediately after it is created if it is allowed to access the wireless channel. However, it is not always good for the sensor to transmit its data packet due to the harsh wireless channel. 

In this paper, we consider the sensor's queue is first-in-first-out. When the packet is generated, it is able to make a decision to transmit that packet or save it in its queue based on its observation of the surrounding environment and the number of packets in its packet queue. If the sensor generates a new packet, but its queue is full, it has to drop the oldest packet. If it sends a packet but cannot receive an acknowledgment, it saves the packet in its queue and it will be retransmitted in next period $T$. Since we consider the sensor communicates directly with the BS, there is not relaying packets.

The queue can save up to $N_q$ packets. The unsuccessful transmission does not drop a packet since the packet is saved and will be retransmitted. Only if the queue is full and the oldest packet is dropped. 
The state transition probability for the queue with $t_{max}$ scheduled transmissions can be written as
\begin{align}
\label{equ:queue}
&P_a(Q_1'=q_1-t_{max}+1+N_u | Q_1=q_1)=\\
&\begin{cases}
& \binom{t_{max}}{N_u}[1-(1-P_e)^{P_L}]^{N_u}(1-P_e)^{P_L(t_{max}-N_u)}\\
&~~~~~~~~~~~~~~~~~\text{ when } 0\leq p \leq t_{max}\\
& 0,~~~~~~~~~~~~~~~ \text{otherwise}
\end{cases}
\end{align}
where $P_L$ is the packet length, $N_u$ is the unsuccessfully transmitted packet number, $P_e$ is the bit-error-rate (BER), and $a$ is the transmission action that is taken at the current state.
\subsection{Wireless Communication Policies}
The action policy consists of choosing the modulation scheme and determining the transmitting packet number. The modulation schemes are MPSK. We consider the allocated time slot number for BPSK is $N_{pmax}=t_{max}$. Using the same time, this can accommodate $N_{pmax}=2t_{max}$ packets for QPSK and $N_{pmax}=3t_{max}$ for 8PSK. As a result, the allocated time interval is a constant, but the maximum number of transmitted packets are different, which depends on the modulation scheme.   
The BER for MPSK modulation can be written as \cite{lu1999m}
\begin{align}
P_{e}\approx \frac{2}{\max (\log_2 M,2)}\sum_{i=1}^{\max (M/4,1)}Q\left(\sqrt{\frac{2P_t l_p}{\eta}}\sin\frac{(2i-1)\pi}{M}\right),
\end{align}
where $Q(x)=(1/\sqrt{2\pi})\int_{x}^{\infty} e^{-t^2/2}dt$ and $M$ can be 2, 4, or 8. Also, we consider the sensor transmission power is a constant. 

\subsection{Rewards Model}
Wireless communication consumes a large portion of a sensor's energy. To efficiently use energy, a sensor prefers to transmit when the channel status is good, while to save the packet in its queue when the path loss is high. Our reward model consists of two key components, namely, the successfully transmitted packet and the unsuccessfully transmitted packet, both of which are scaled by the overall transmission energy. In general, the reward is defined as 
\begin{align}
\label{equ:reward}
{\mathcal R}_a(s) =   \frac{\left[N_t-\alpha_1 N_u-\alpha_2(q_1-N_t+1)\right]\log_2(M)}{t_{sym}P_{t}(N_t+N_u)}
\end{align}
where $N_t$ is the successfully transmitted packet number, $t_{sym}$ the symbol time, $\alpha_1$ is the coefficient to scale dropped packets number, and $\alpha_2$ is the coefficient to scale the packet number in the queue. If $N_t>N_u$, the sensor receives a positive reward. If $N_t = N_u$ or there is no packet being transmitted, the reward is 0. If $N_t<N_u$, the sensor transmits when BER is large and it receives negative rewards. 

The $\alpha_1 N_u$ in \eqref{equ:reward} penalizes the unsuccessfully transmitted packets. In this way, if the channel path loss is high, the sensor can try to use low level modulation. Although the unsuccessfully transmitted packets have been counted in the denominator, the $\alpha_1 N_u$ provides more flexibility. If $\alpha_1$ is large, the sensor is more cautious to transmit a packet and more likely to save the packet in its queue.  
By using $\alpha_2(q_1-N_t+1)$, the sensor tends to send more packets to reduce the number of packets in its queue. In other words, when the channel path loss is small, the more transmitted packets, the more rewards. This is also equivalent to control the delay, i.e., a long queue causes more packets being delayed. 
\section{Optimal Policy using Reinforcement Learning}
Based on the developed state, action, and reward model, we provide the RL algorithm in this section. The Bellman's equation for the expected value function can be written as
\begin{align}
V_{\pi^{*}}(s)=\max _{a \in \mathcal{A}}\left({\mathcal R}_{a}(s)+\lambda \sum_{s' \in \mathcal{S}} P_{a}\left(s' | s\right) V_{\pi^{*}}\left(s'\right)\right)
\end{align} 
where $V_{\pi^{*}}(s)$ is value function at state $s$ using the optimal policy $\pi^{*}$, $a \in \mathcal{A}$ is one of the possible actions, $s'$ is the next state,  $P_{a}\left(s' | s\right)$ is the transition probability from $s$ to $s'$ using action $a$, $\mathcal{S}$ contains all the states, and $\lambda$ is a discount scalar. The transition probability can be expressed as
\begin{align}
P_{a}\left(s' | s\right) = P(c'|c)\cdot P_a(Q_1'|Q_1)
\end{align} 
where $P(c'|c)$ is trained from the data and is shown in Fig. \ref{fig:transition_prob},  and $P(Q_1'|Q_1)$ is the queue transition probability, which is given in \eqref{equ:queue}. Then, the value iteration can be employed to obtain the optimal policy \cite{sutton2018reinforcement}, i.e.,
\begin{align}
&V_{k+1}^a(c,q_1) = {\mathcal R}_a(c,q_1)+\lambda \sum_{(c',q_1') \in \mathcal{S}}P_a (c',q_1'|c,q_1)V_k (c',q_1')\\
&V_{k+1}(c,q_1) = \max_{a \in \mathcal{A}} V_{k+1}^a(c,q_1)
\end{align} 
where the subscript $k$ denotes the iteration round. 

To gain more insights, we focus on the value function and the reward model to see how the modulation scheme, queue length, and the environmental change affect each other and we try to understand how the wireless sensor responds to the change.

Intuitively, there is a tradeoff between the delay and the number of dropped packets. If the queue is short, more packets are dropped which reduces the reward, while if the queue is long, the delay increases which also reduces the reward.  However, this is not true since the performance is not only determined by the queue but also the channel. Next, we show that, when $q_1\leq N_{pmax}$, where $N_{pmax}$ is the maximum number of packets that can be transmitted in interval $T$, $V_{\pi^{*}}(s)$ is a constant, while when $q_1>N_{pmax}$, $V_{\pi^{*}}(s)$ is an monotonically decreasing function. Here, we assume the path loss is small which results in a small BER. Consider the values functions $V_{k+1}^a(c,q_1)$ and $V_{k+1}^{a'}(c,q_1+1)$, the latter has one more packet in its queue than the former and they have different actions, i.e., $a$ and $a'$. First, we assume the queue length is small and the channel does not change. To compare the values, we have
\begin{align}
&V_{k+1}^a(c,q_1)-V_{k+1}^{a'}(c,q_1+1)={\mathcal R}_a(c,q_1)-{\mathcal R}_{a'}(c,q_1+1)\nonumber \\
&\label{equ:value1}+\lambda P_a(c,q_1'|c,q_1)V_k(c,q_1')-\lambda P_{a'}(c,q_1^{\prime\prime}|c,q_1+1)V_k(c,q_1^{\prime\prime}).
\end{align} 
Since we consider $q_1$ is small and the channel status is good, the sensor can empty its queue with probability 1. Thus, $q_1'=q_1^{\prime\prime}$ and $\lambda P_a(c,q_1'|c,q_1)V_k(c,q_1')-\lambda P_{a'}(c,q_1^{\prime\prime}|c,q_1+1)V_k(c,q_1^{\prime\prime})=0$. Also, since BER is small, $N_u=0$, and $q_1-N_t+1=0$, \eqref{equ:reward} can be simplified to ${\mathcal R}_a(c,q_1)={\mathcal R}_{a'}(c,q_1+1)=\log_2(M)/(t_{sym}P_t)$. Thus, when $q_1\leq N_{pmax}$, \eqref{equ:value1} is equivalent to 0 and the value functions are constants, which are not affected by the queue length. 

When $q_1> N_{pmax}$, the sensor cannot transmit all the packets in its queue and there are some packets left after transmission. In this case, the longer the queue, the smaller the reward. In \eqref{equ:value1}, ${\mathcal R}_a(c,q_1)$ is larger than ${\mathcal R}_{a'}(c,q_1+1)$ since $q_1-N_t+1<q_1-N_t+2$. Then, we can apply the induction method. When $q_1'$ and $q_1^{\prime\prime}$ are smaller than $N_{pmax}$, $V_k(c,q_1')=V_k(c,q_1^{\prime\prime})$. Since ${\mathcal R}_a(c,q_1)$ is larger, $V_{k+1}^a(c,q_1)-V_{k+1}^{a'}(c,q_1+1)>0$. When, $q_1'=N_{pmax}$ and $q_1^{\prime\prime}=N_{pmax}+1$, $V_k(c,q_1')>V_k(c,q_1^{\prime\prime})$ which can be proved by using the conclusion when $q_1'$ and $q_1^{\prime\prime}$ are smaller than $N_{pmax}$. As a result, $V_{k+1}^a(c,q_1)-V_{k+1}^{a'}(c,q_1+1)>0$, which is still valid. Similarly, we can prove that it is valid when $q_1'$ and $q_1^{\prime\prime}$ both are larger than $N_{pmax}$. Here, we find that a long queue cannot guarantee a larger expected value, since it creates larger delay. When path loss is small, the long queue does not help and the optimal strategy is sense-then-transmit. If the packet number in the queue is larger than $N_{pmax}$, the expected value reduces. 

When the channel path loss is high, the state transition probability is not 1 and $N_u$ can be larger than 0. In this case, when $q_1<N_{pmax}$ the expected value is not a constant since neither the rewards or value functions are equivalent. With more transmission errors, the long queue can provide more space to save packets to avoid unsuccessful transmissions with high path loss. Also, depends on the tolerance of delay, we can adjust $\alpha_2$ to provide more flexibility.     

\section{Numerical Simulations}

In this section, we numerically evaluate the proposed solution. Here, we compare with two baseline models, the sense-then-transmit using BPSK modulation and 8PSK modulation. Note that, there is no queue for the baseline models since we consider the traditional wireless underground sensor communication model without considering dynamic environmental change. The simulation parameters are given in Table \ref{tab:parameter}. 
\begin{table}[t]
	\renewcommand{\arraystretch}{1.3}
	\caption{Simulation Parameters}
	\label{tab:parameter}
	\centering
	\begin{tabular}{c|c||c|c}
		\hline
		Symbol  &  Value &Symbol  &  Value\\
		\hline
		\hline
		
		$\eta$   & -100~dBm& $N_{q}$ &150   \\
		\hline
		
		$t_{max}$ & 15  &$t_{sym}$&1/60000\\
		\hline
			$P_L$ & 1000  &$\alpha_2$&0.1\\
		\hline
			$\alpha_1$ & 1  &$\lambda$&0.1\\
		\hline
	\end{tabular}
\end{table}

\begin{figure}[t]
	\centering
	\includegraphics[width=0.35\textwidth]{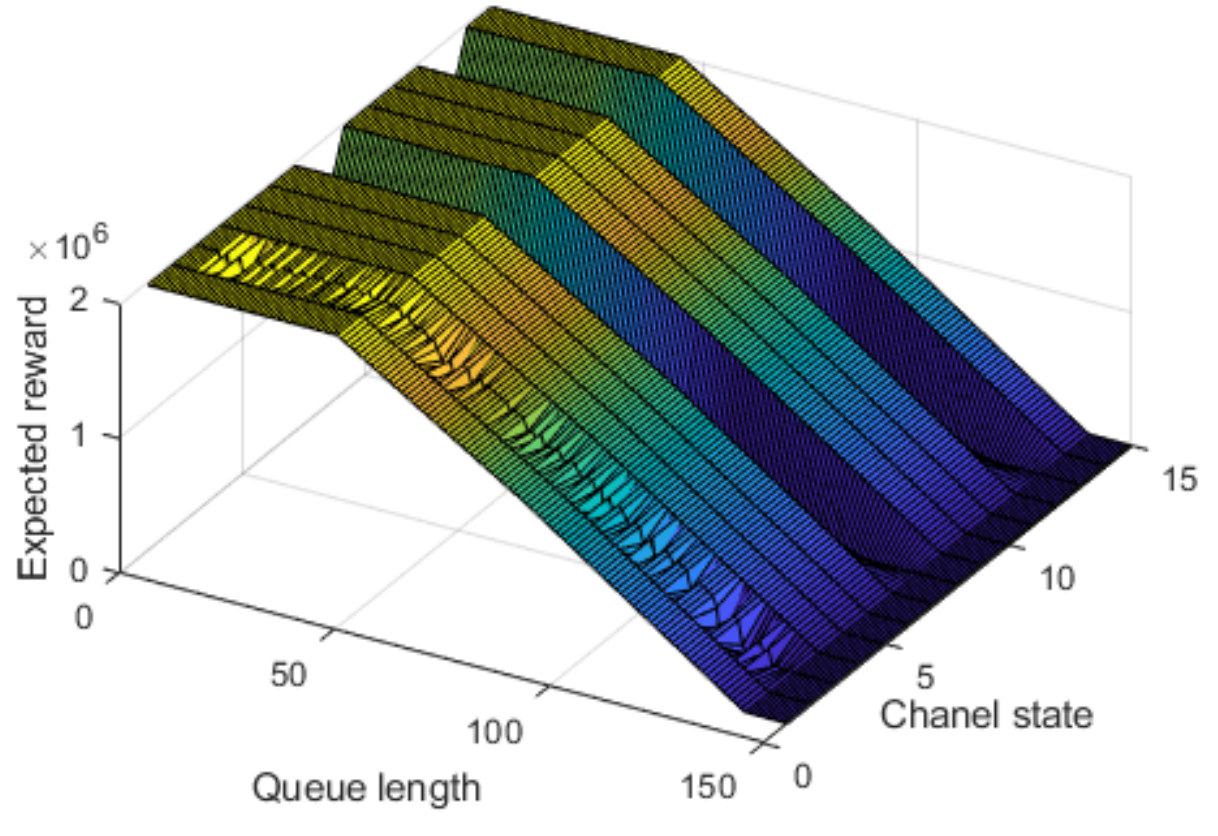}
	\vspace{-5pt}
	\caption{Expected reward at different states.}
	\vspace{-10pt}
	\label{fig:reward}
\end{figure}

First, we show how the expected rewards change at different states. The state is composed of queue length and channel state. The transmission power is 20 dBm to provide small BER. Note $N_{pmax}= 3t_{max}$ since the 8PSK is used when the signal-to-noise ratio (SNR) is high. As shown in Fig. \ref{fig:reward}, when the queue length is smaller than $N_{pmax}$, the expected reward is a constant, while when the queue length becomes larger, the expected reward decreases. This shows that when SNR is high, i.e., equivalently the path loss is small, there is no need to use long queues since the expected reward is no larger than that with one packet in the queue. This is consistent with the discussion in the previous section.  

\begin{figure}[t]
	\centering
	\includegraphics[width=0.35\textwidth]{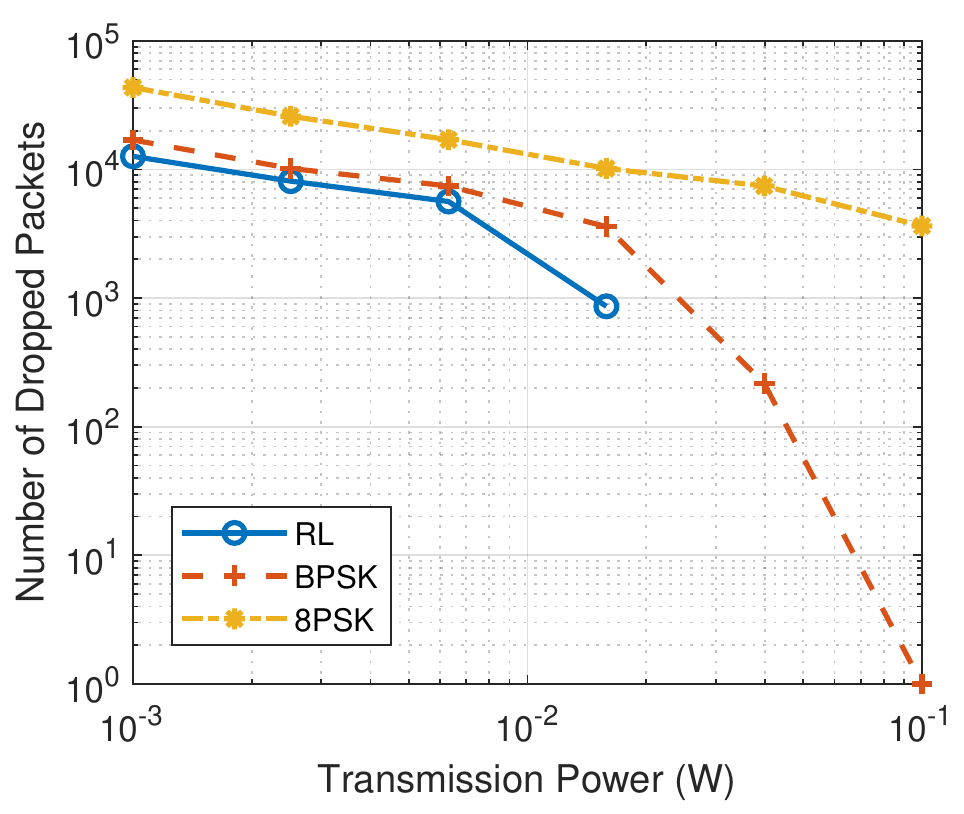}
	\vspace{-5pt}
	\caption{The number of dropped packets.}
	\vspace{-10pt}
	\label{fig:dropped}
\end{figure}

The number of dropped packets is shown in Fig.~\ref{fig:dropped}. The transmission power is increased from 0.001 W to 0.1 W. The sense-then-transmit using BPSK and 8PSK do not have queues. The 8PSK suffers from higher BER, and thus its dropped packet number is large. The BPSK is more reliable. However, there are some extreme scenarios where the path loss is very high. Even using the BPSK modulation, we cannot avoid dropping packets. We notice that the RL solution can significantly reduce the number of dropped packets, e.g., when the transmission power is slightly higher than 0.01 W, the dropped packets number becomes 0. The RL solution can observe the environment and save packets when the channel has high path loss. It waits until the channel becomes better to transmit the packets. Therefore, it enjoys smaller dropped packets number. One may argue that the benefits of using RL arises from the queue. However, even we provide queues to the baseline models, it requires a strategy to transmit the packets in the queue. Depending on the intelligence level of the transmission strategy, the performances can be drastically different. Currently, there is no existing research efforts on this topic and we only consider the simple baseline models.

\begin{figure}[t]
	\centering
	\includegraphics[width=0.35\textwidth]{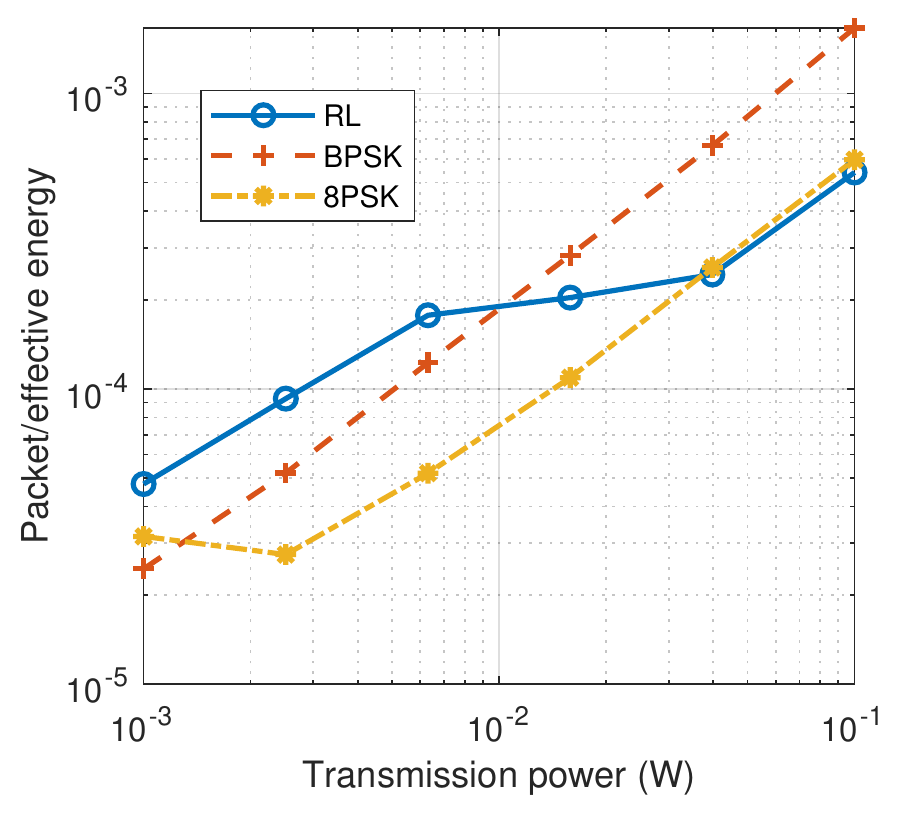}
	\vspace{-5pt}
	\caption{Effective energy used to successfully transmit a packet.}
	\vspace{-10pt}
	\label{fig:energy}
\end{figure}

In Fig.~\ref{fig:energy}, we show the ratio of the overall energy used over the successfully transmitted packet number, i.e., $[{t_{sym}P_{t}(N_t+N_u)}]/N_t$. Note that, the smaller the ratio, the better the energy is used since a unit energy can successfully transmit more packets. From the figure, we can see that with low transmission power, RL solution has slightly higher ratio. Because the RL solution prefers to not transmit packets when transmission power is low and the channel is not good. If the transmission fails, it saves the packet and retransmit it. This increases the overall transmission number since the low SNR causes most of the transmissions failing. As the transmission power increases, the SNR becomes large and the RL solution can successfully transmit more packets. Moreover, since the SNR is large, the RL solution uses 8PSK modulation and therefore its performance converges to the 8PSK model. 
\begin{figure}[t]
	\centering
	\includegraphics[width=0.35\textwidth]{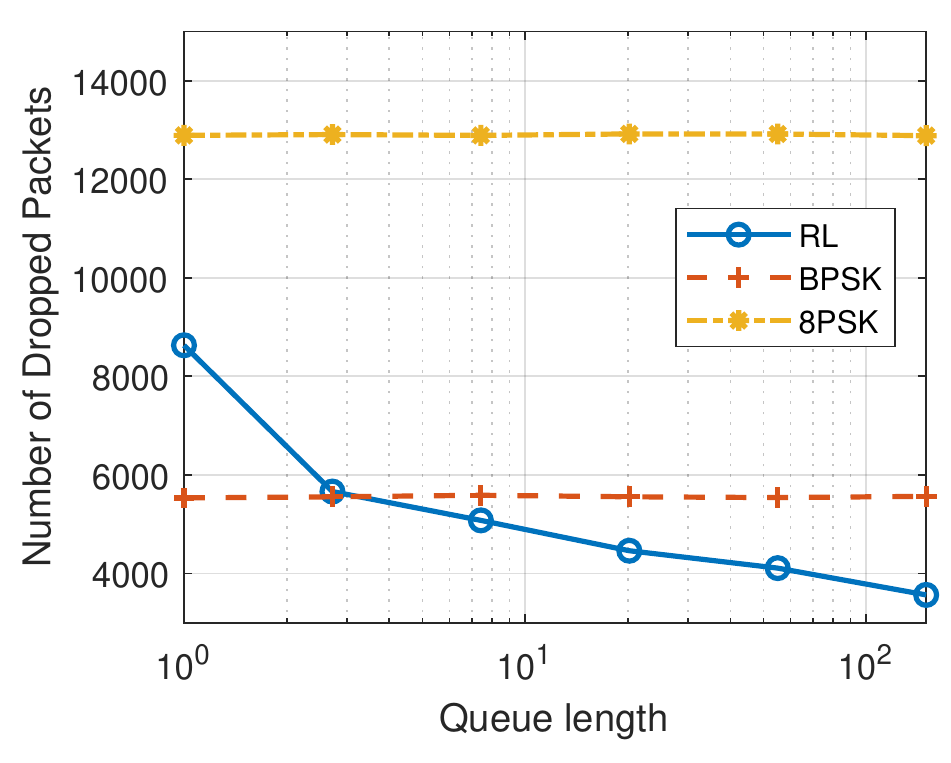}
	\vspace{-5pt}
	\caption{Effect of queue length of the number of dropped packets.}
	\vspace{-10pt}
	\label{fig:queue}
\end{figure}

In Fig.~\ref{fig:queue}, we show the effect of $N_q$. The transmission power is 0.01 W. The maximum transmit packet number in an interval $T$ is $t_{max}=\lceil 0.1 N_q \rceil$. As we can see in the figure, when $N_q$ is small, the number of dropped packets is large, since the SNR is not high enough to avoid errors and the queue is not long enough to save packets. As the queue length becomes longer, the dropped packets number reduces and it is even smaller than the BPSK baseline model.  
\begin{figure}[t]
	\centering
	\includegraphics[width=0.4\textwidth]{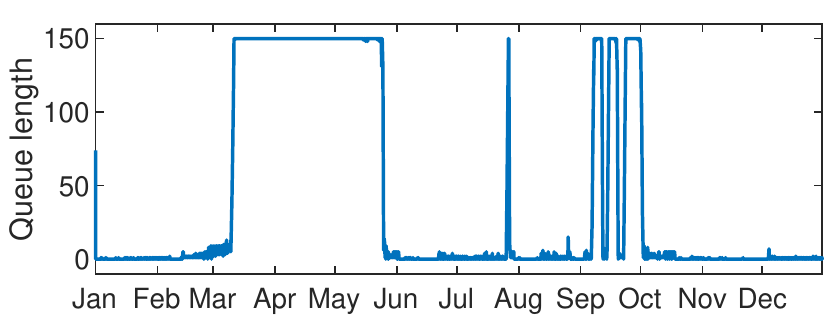}
	\vspace{-5pt}
	\caption{The status of queue length.}
	\vspace{-10pt}
	\label{fig:queueL}
\end{figure}

In Fig.~\ref{fig:queueL}, we show an example of the packet number in the queue. The transmission power is 0.001 W and all other parameters are given in Table~\ref{tab:parameter}. The results suggest that when the path loss is high, the sensor tries to save packets and only transmit the minimum number of packet, i.e., one packet, to accommodate the new packet. Refer back to Fig.~\ref{fig:pathloss}, the path loss from March to June and from September to October are large and that is when the sensor's queue is almost full. 
\section{Conclusion}
The dynamic change of dielectric parameters in underground environments poses significant challenges in designing reliable wireless sensor networks. The network performance can easily be deteriorated by the dynamic change. In this paper, we propose a data-driven model for adaptive wireless communication in underground using reinforcement learning. We provide a systematic model to capture the change of dielectric parameters and relate the change to wireless channels. We consider the sensor has a queue, which can be leveraged to avoid unsuccessful transmissions. The optimal communication policies are derived and evaluated. The results show that the proposed solution can significantly reduce the number of dropped packets with a controllable delay and reasonable energy consumption. Our future work will consider not only the underground environmental change but also the RF environmental change such as interference and noise, as well as the sensor's battery. Also, instead of using a model-based solution, we will investigate the model-free solution using Q-learning to make the system more adaptive. 
\bibliographystyle{IEEEtran}
\bibliography{guo}
\end{document}